\documentclass[runningheads]{ceurart}
\usepackage[T1]{fontenc}
\usepackage{graphicx}
\usepackage{hyperref}
\usepackage{booktabs}
\usepackage{multicol}
\usepackage{multirow}

\begin{document}
\title{\includegraphics[scale=0.02]{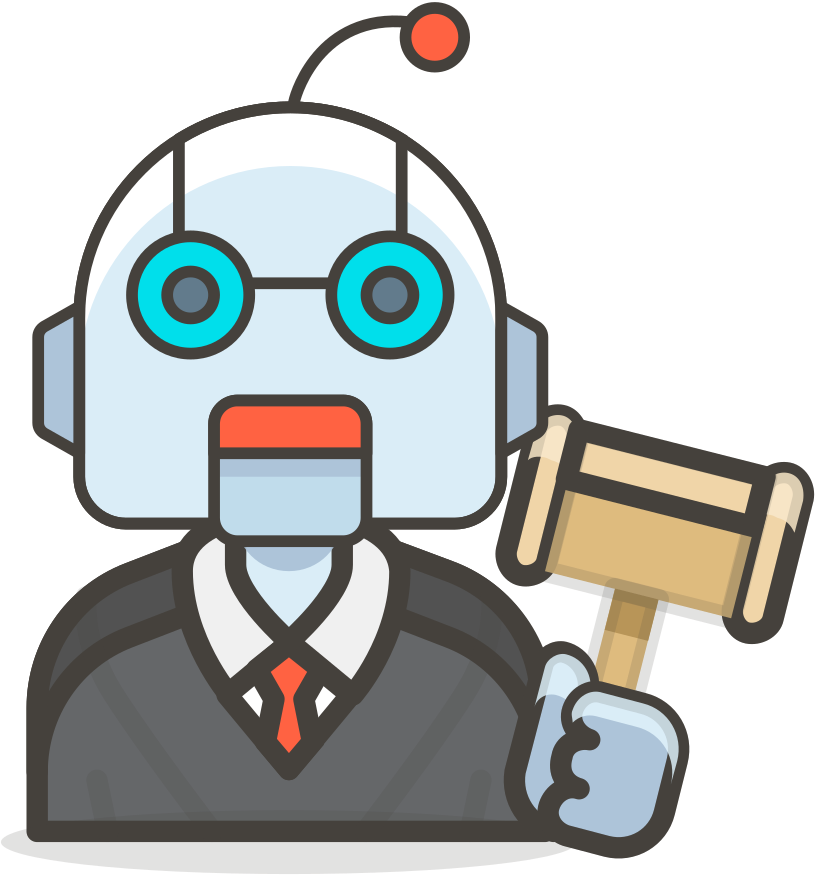} LLMJudge: LLMs for Relevance Judgments}
%
%
\author[1]{{Hossein A.} Rahmani}
\address[1]{University College London, London, UK}

\author[1]{Emine Yilmaz}

\author[2]{Nick Craswell}
\address[2]{Microsoft, Seattle, US}

\author[3]{Bhaskar Mitra}
\address[3]{Microsoft, Montréal, Canada}

\author[4]{Paul Thomas}
\address[4]{Microsoft, Adelaide, Australia}

\author[5]{{Charles L. A.} Clarke}
\address[5]{University of Waterloo, Ontario, Canada}

\author[6]{Mohammad Aliannejadi}
\address[6]{University of Amsterdam, Amsterdam, The Netherlands}

\author[6]{Clemencia Siro}

\author[7]{Guglielmo Faggioli}
\address[7]{University of Padua, Padua, Italy}

%
%
%

\copyrightyear{2024}
\copyrightclause{Copyright for this paper by its authors.
  Use permitted under Creative Commons License Attribution 4.0
  International (CC BY 4.0).}

\conference{LLM4Eval: The First Workshop on Large Language Models for Evaluation in Information Retrieval, 18 July 2024, Washington DC, United States}

\maketitle              
%

%
%


\section{Introduction}
The \texttt{LLMJudge} challenge\footnote{\url{https://llm4eval.github.io/challenge/}} is organized as part of the \texttt{LLM4Eval}\footnote{\url{https://llm4eval.github.io/}} workshop \cite{rahmani2024llm4eval} at SIGIR 2024. Test collections are essential for evaluating information retrieval (IR) systems. The evaluation and tuning of a search system is largely based on relevance labels, which indicate whether a document is useful for a specific search and user. However, collecting relevance judgments on a large scale is costly and resource-intensive. Consequently, typical experiments rely on third-party labelers who may not always produce accurate annotations. The \texttt{LLMJudge} challenge aims to explore an alternative approach by using LLMs to generate relevance judgments. Recent studies have shown that LLMs can generate reliable relevance judgments for search systems. However, it remains unclear which LLMs can match the accuracy of human labelers, which prompts are most effective, how fine-tuned open-source LLMs compare to closed-source LLMs like GPT-4, whether there are biases in synthetically generated data, and if data leakage affects the quality of generated labels. This challenge will investigate these questions, and the collected data will be released as a package to support automatic relevance judgment research in information retrieval and search.


\section{Related Work}
Automatic relevance judgment has recently received significant attention in the Information Retrieval (IR) community. In earlier studies, Faggioli et al.~\cite{faggioli2023perspectives} studied different levels of human and LLMs collaboration for automatic relevance judgement. They suggested the need for humans to support and collaborate with LLMs for a human-machine collaboration judgment. Thomas et al.~\cite{thomas2023large} leverage LLMs capabilities in judgement at scale, in Microsoft Bing. They used real searcher feedback to consider an LLM and prompt in a way that matches the small sample of searcher preferences. Their experiments show that LLMs can be as good as human annotators in indicating the best systems. They also comprehensively investigated various prompts and prompt features for the task and revealed that LLM performance on judgments can varies with simple paraphrases of prompts. Recently, Rahmani et al.~\cite{rahmani2024synthetic} have studied fully synthetic test collection using LLMs. In their study, they not only generated synthetic queries but also synthetic judgment to build a full synthetic test collation for retrieval evaluation. They have shown that LLMs are able to generate a synthetic test collection that results in system ordering performance similar to evaluation results obtained using the real test collection.

\section{LLMJudge Task Design}
The challenge will be, given the query and document as input, how they are relevant. Here, we use four-point scale judgments to evaluate the relevance of the query to document as follows:

\begin{itemize}
\item\textbf{[3] Perfectly relevant}: The passage is dedicated to the query and contains the exact answer. 
\item\textbf{[2] Highly relevant}: The passage has some answers for the query, but the answer may be a bit unclear, or hidden amongst extraneous information. 
\item\textbf{[1] Related}: The passage seems related to the query but does not answer it. 
\item\textbf{[0] Irrelevant}: The passage has nothing to do with the query. 
\end{itemize}

The task is, by providing the datasets that include queries, documents, and query-document files to participants, to ask LLMs to generate a score [0, 1, 2, 3] indicating the relevance of the query to the document.

\section{LLMJudge Data}
The \texttt{LLMJudge} challenge dataset is built upon the passage retrieval task dataset of the
TREC 2023 Deep Learning track\footnote{\url{https://microsoft.github.io/msmarco/TREC-Deep-Learning.html}} (TREC-DL 2023) \cite{craswell2024overview}. Table \ref{tbl:llmjudge-dataset} shows the statistics of the \texttt{LLMJudge} challenge datasets. We divide the data into development and test sets. The test set is used for the generation of judgment by participants, while the development set could be used for few-shot or fine-tuning purposes. The datasets, sample prompt, and the quick starter for automatic judgment can be found at the following repository: \url{https://github.com/llm4eval/LLMJudge}

\begin{table}
    \caption{Statistics of LLMJudge Dataset}
    \label{tbl:llmjudge-dataset}
        \begin{tabular}{lcc}
            \toprule
            \textbf{} & \textbf{Dev} & \textbf{Test} \\
            \midrule
             \# queries  & 25    & 25 \\
             \# passage & 7,224 & 4,414 \\
             \# qrels    & 7,263 & 4,423 \\
             \midrule
             \# irrelevant (0)         & 4,538 & 2,005 \\
             \# related (1)            & 1,403 & 1,233 \\
             \# highly relevant (2)    & 625 & 808 \\
             \# perfectly relevant (3) & 697 & 377 \\
            \bottomrule
        \end{tabular}
\end{table}

\section{Evaluation}
Participants’ results will then be evaluated in two methods after submission:
\begin{itemize}
    \item automated evaluation metrics on human labels in the test set hidden from the participants;
    \item system ordering evaluation of multiple search systems on human judgments and LLM-based judgments
\end{itemize}

\section{Submissions and Results}
In order to evaluate the quality of the generated labels, we used Cohen's $\kappa$ to see the labeler's agreement with LLMJudge test data at query-document level and the Kendall's $\kappa$ to check the labeler's agreement with LLMJudge test data on system ordering, i.e., the runs that submitted to TREC DL 2023. In total, we had 39 submissions (i.e., the 39 labelers) from $7$ groups from National Institute of Standards and Technology (NIST), RMIT University, The University of Melbourne, University of New Hampshire, University of Waterloo, Included Health, and University
of Amsterdam.

\begin{figure}
    \centering
    \includegraphics[width=0.5\linewidth]{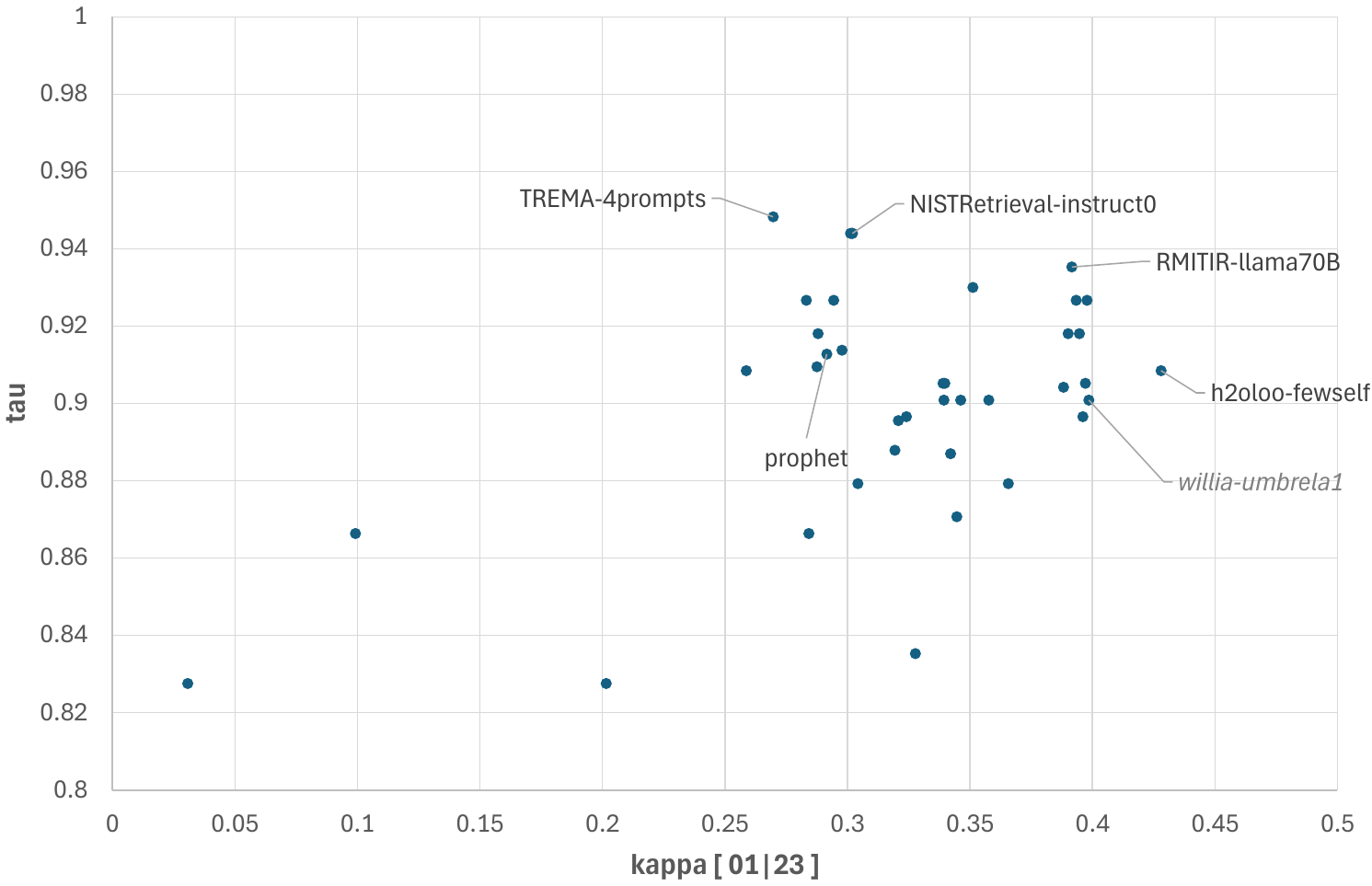}
    \caption{Scatter plot of Cohen's $\kappa$ and Kendall's $\tau$ for submitted labelers}
    \label{fig:llmjudge-result}
\end{figure}

Figure \ref{fig:llmjudge-result} shows the performance of submitted labelers on the \texttt{LLMJudge} test set. The x-axis represents Cohen's $\kappa$, and the y-axis shows the labelers' agreement on system ordering. Labelers exhibit low variability in Kendall's $\tau$ but greater variability in Cohen's $\kappa$. Most labelers cluster within a narrow range of $\tau$ values, indicating consistent system rankings but more variation in inter-rater reliability, as measured by Cohen's $\kappa$. This suggests that while labelers generally agree on rankings, their exact labels are less consistent, leading to the observed variability in $\kappa$.

\section*{Acknowledgment}
The challenge is organized as a joint effort by the University College London, Microsoft, the University of Amsterdam, the University of Waterloo, and the University of Padua. The views expressed in the content are solely those of the authors and do not necessarily reflect the views or endorsements of their employers and/or sponsors. This work is supported by the Engineering and Physical Sciences Research Council [EP/S021566/1], the EPSRC Fellowship titled ``Task Based Information Retrieval'' [EP/P024289/1], CAMEO, PRIN 2022 n. 2022ZLL7MW. 

%
%
%
\bibliography{mybibliography}
\end{document}